\documentclass[aps,pre,showpacs,superscriptaddress,reprint]{revtex4-1}
\usepackage[english]{babel}
\usepackage[pdftex]{graphicx} 
\usepackage{bm} 			  
\usepackage{hyperref} 		  
\usepackage{amsmath}
\usepackage{amstext}
\usepackage{natbib}

\usepackage{color}
\newcommand{\cg}{\color{black}}

\begin{document}

\title{  
    From Network Reliability to the Ising Model:  \\
A Parallel Scheme for Estimating the Joint Density of States
}

\author{Yihui Ren}
\email[]{yren2@vbi.vt.edu}
\affiliation{Network Dynamics and Simulation Science Laboratory, Biocomplexity
Institute of Virginia Tech, Blacksburg, VA, 24061, USA}

\author{Stephen Eubank}
\email[]{seubank@vbi.vt.edu}
\affiliation{Network Dynamics and Simulation Science Laboratory, Biocomplexity
Institute of Virginia Tech, Blacksburg, VA, 24061, USA}
\affiliation{Department of Physics, Virginia Tech, Blacksburg, Virginia 24061, USA}
\affiliation{Department of Population Health Sciences, Virginia Tech, Blacksburg, Virginia 24061, USA}

\author{Madhurima Nath} 
\email[]{nmaddy@vbi.vt.edu}
\affiliation{Network Dynamics and Simulation Science Laboratory, Biocomplexity
Institute of Virginia Tech, Blacksburg, VA, 24061, USA}
\affiliation{Department of Physics, Virginia Tech, Blacksburg, Virginia 24061, USA}

\date{\today}

\begin{abstract} 
    Network reliability is the probability that a dynamical system composed of
    discrete elements interacting on a network will be found in a configuration
    that satisfies a particular property.  We introduce a new reliability
    property, Ising-feasibility, for which the network reliability is the Ising
    model's partition function.  As shown by Moore and Shannon, the network
    reliability can be separated into two factors: \textit{structural}, solely
    determined by the network topology, and \textit{dynamical}, determined by
    the underlying dynamics.  In this case, the structural factor is known as
    the joint density of states.  Using methods developed to approximate the
    structural factor for other reliability properties, we simulate the joint
    density of states, yielding an approximation for the partition function.
    Based on a detailed examination of why na\"ive Monte Carlo sampling
    gives a poor approximation, we introduce a novel parallel scheme for
    estimating the joint density of states using a Markov chain Monte Carlo method
    with a spin-exchange random walk. This parallel scheme makes simulating the
    Ising model in the presence of an external field practical on small
    computer clusters for networks with arbitrary topology with $\sim 10^6$
    energy levels and more than $10^{308}$ microstates.
\end{abstract}


\maketitle

\section{Introduction}

The Ising model~\cite{Lenz1920,Ising1925} of ferromagnetism in crystals has
been the object of sustained scrutiny since its introduction nearly a century
ago, due to the rich phenomenology it produces from simple
dynamics~\cite{McCoyMaillard2012,Taroni2015}.  The Ising model has also had a
far-reaching influence in domains ranging from protein
folding~\cite{TanakaScheraga1977} to social
science~\cite{Schelling1971,Stauffer2007}.  Yet it has proven resistant to
analytical solution, except in special cases such as 1 or 2-dimensional
lattices with no external field.  Indeed, solving the model in the general case
is known to be NP-hard~\cite{Barahona1982,UngerMoult1993,Istrail2000}.  Hence
we largely depend on approximations or numerical simulations for understanding
its properties.  Unfortunately, the na\"ive Metropolis algorithm suffers from
poor convergence at precisely the most interesting region of parameter space,
the critical point~\cite{Metropolis1953,Swendsen1987}.  Wang and
Landau~\cite{Wang2001,Wang2001a} proposed a more efficient algorithm that
focuses on estimating the density of states.  Once the density of states is
known, the system's partition function and related thermodynamic quantities can
be computed without further simulation.  The original sequential Wang-Landau
method is not practical for large systems, because its convergence time
increases rapidly with the number of energy states.  The state-of-the-art
replica-exchange framework~\cite{Vogel2013,Vogel2014} provides a parallel
algorithm to estimate the univariate density of states $g\left( k \right)$.
However, it is not clear how to apply this parallel scheme to estimate the
joint density of states $g\left( k,v \right)$, necessary for computing physical
quantities in the presence of an external field.  Here we use insights from
Moore-Shannon network reliability~\cite{Moore1956191,Moore1956281} to construct
a new parallel scheme that bridges the gap between the Wang-Landau approach and
the estimation of joint density of states $g\left( k,v \right)$. The result is
an efficient estimation scheme for the partition function of the Ising model in
the presence of an external field that performs well even on large, irregular
networks.

The Ising model is defined on a graph $G\left(V,E\right)$ with vertex and edge
sets $V$ and $E$, respectively, by the Hamiltonian 
\begin{equation}
H=J\sum_{\left(i,j\right)\in E}\sigma_{i}\sigma_{j}+\mu B\sum_{i\in V}\sigma_{i},\label{eq:Hamiltonian}
\end{equation}
where $\sigma_{i}\in\left\{ -1,1\right\} $ represents the state of the vertex
$i$. $J$ is the coupling strength between neighboring vertices and $B$ is the
external field. The exact solution of the Ising model in one dimension does not
exhibit any critical phenomena.  In the study of the order-disorder
transformation in alloys, Bragg and Williams~\cite{Bragg1934,Bragg1935} used a
mean-field approximation for the Hamiltonian in which each individual vertex
interacts with the mean state of the entire system.  This is known as the
Bragg-Williams approximation or the zeroth approximation of the Ising
model~\cite{Pathria1996Book}. An analytic expression for the partition function
of a two-dimensional Ising model \textit{in the absence of an external field}
was given by Onsager~\cite{Onsager1944} and later derived rigorously by C.\ N.\
Yang~\cite{Yang1952}. In spite of great effort in the seven decades since, the
exact solution of the 2D Ising model in the presence of an external field
remains unknown. 

A network's reliability is the probability it ``functions'' -- i.e., continues
to have a certain structural property -- even under random failures of its
components.  It was proposed in 1956 by Moore and
Shannon~\cite{Moore1956191,Moore1956281} as a theoretical framework for
analyzing the trade-off between reliability and redundancy in telephone relay
networks.  The desired structural property in that case, known as
``two-terminal'' reliability,  is to have a communication path between a
specified source node and specified target node.  Since then, a wide variety of
properties have been studied, for example: ``all-terminal'' reliability
requires the entire graph to be connected; ``attack-rate-$\alpha$'' reliability
requires the root-mean-square of component sizes is no less than $\alpha
N$~\cite{Youssef2013}.  Network reliability can be expressed as a polynomial in
parameters of the dynamical system whose coefficients encode the interaction
network's structure.  A reliability polynomial is the partition function of a
physical system~\cite{EssamTsallis86,welsh2000potts,beaudin2010little}, but it
emphasizes the role of an interaction network's structure rather than the form
of the interactions.

Specifically, the reliability of an interaction network $G\left(V,E\right)$ is
\begin{equation}
    R\left(x;r,G\right)\equiv\sum_{s\in\mathcal{S}}r\left(s\right)p_{s}\left(x\right),
\end{equation} where $\mathcal{S}$ is the set of all subgraphs of
$G\left(V,E\right)$; $r\left(s\right)\in\left\{0,1\right\}$ is a binary
function indicating whether the subgraph $s$ has the desired property, i.e.
``two-terminal''; and $p_{s}\left(x\right)$ is the probability resulting in a
modified interaction subgraph $s$.  The probability of picking a subgraph
$p_{s}\left(x\right)$ reflects random, independent edge failures in the network
with a failure rate $\left( 1-x
\right)\in\left[0,1\right]$~\cite{Moore1956191}.  Hence, with
$M\equiv\left|E\right|$, $p_{s}\left(x\right)=x^{k}\left(1-x\right)^{M-k}$
where $k$ is the number of edges in the subgraph $s$.  

If we group all $2^{M}$ subgraphs into $M$ equivalence classes by the number of
edges in the subgraphs, the reliability can be expressed as \begin{equation}
    R\left(x;r,G\right)=\sum_{k=1}^{M}R_{k}\left( r,G
    \right)x^{k}\left(1-x\right)^{M-k} \label{eq:Rk} \end{equation} where
$R_{k}$ is the number of subgraphs with $k$ edges that have the desired
property.  As shown in Section~\ref{sec:ReliParty}, $R_k$ is equivalent to the
density of states in the Ising model.

Evaluating $R_k$ exactly is known to be as difficult as $\#P-complete$.
In practice, however, $R_k$ can be estimated by $R_{k}=P_{k}{M \choose k}$, 
where $P_{k}$ is the fraction of subgraphs with the desired property, which
can be estimated via sampling. 

In summary, the reliability of a graph $G\left(V,E\right)$ with respect to a
certain binary criterion can be written as a polynomial:
\begin{equation}
R\left(x;r,G\right) = \sum_{k=1}^{M}P_{k}\left( r,G \right){M 
\choose k}x^{k}\left(1-x\right)^{M-k} \label{eq:RxrG}
\end{equation}
Each term in the reliability polynomial, Eq.~(\ref{eq:RxrG}), contains two
independent factors: a \textit{structural} factor $P_{k}$ and a
\textit{dynamical} factor $x^{k}\left(1-x\right)^{M-k}$.  The reason for
calling this factor ``dynamical'' will become apparent in
Section~\ref{sec:deltaAppr}.  The structural factor depends only on the
topology of the graph $G$ and the reliability criterion $r$, whereas the
dynamical factor only depends on the parameter $x$ -- for given values of
$P_k$, the reliability $R$ is a function of $x$ alone.

This separation of \textit{dynamical} and \textit{structural} factors suggests
new, more efficient ways to simulate Ising models. In
Section~\ref{sec:ReliParty}, we will illustrate that the reliability $R\left( x
\right)$ is equivalent to the partition function $Z\left( \beta \right)$ of the
Ising model; and the ``failure rate'' $1-x$ actually corresponds to physical
quantities such as the temperature, the external field and the coupling
strength in the Ising model.  In Section~\ref{sec:deltaAppr}, we use this
perspective to show that the Bragg-Williams approximation is given by the
first-order term in a principled approximation to the structural factor.  In
Section~\ref{sec:joint}, we use this perspective to extend the Wang-Landau
method into an efficient parallel scheme for estimating the joint density of
states, which we demonstrate on a $32\times 32$ square lattice and a Cayley
tree.


\section{Network Reliability and Partition Function}\label{sec:ReliParty}

\begin{figure}
\includegraphics[width=0.6\columnwidth]{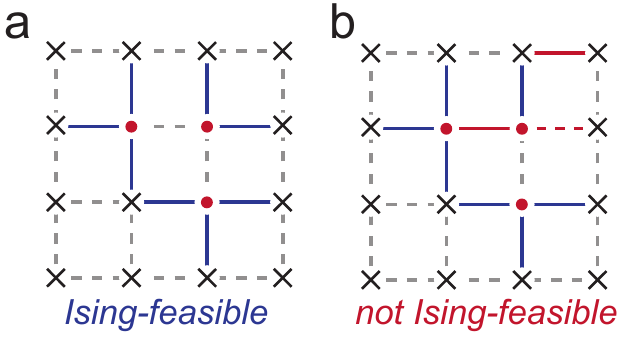}
\protect\caption{\label{fig:IsFeasible}(a) An Ising-feasible configuration with
three spin-up vertices (red dots) and eight discordant spin pairs or edges
(solid line segments). (b) A configuration that is not Ising-feasible because
of the inconsistent edges (red line segments). Independently choosing edges and 
spin-up vertices will rarely produce an Ising-feasible configuration, but any set
of randomly chosen spin-up vertices uniquely determines a set of edges.
}
\end{figure}

The Ising model assumes that the state of a site is binary, either ``spin-down"
($\sigma_{i}=-1$) or ``spin-up" ($\sigma_{i}=1$), and that each site interacts
only with its nearest neighbors, with a coupling strength $J$. All sites are
exposed to a uniform external field $B$.  The collection of all the sites'
states is called a ``microstate'' of the system.  The Hamiltonian for the Ising
model on a graph $G\left(V,E\right)$ is shown in Eq.~(\ref{eq:Hamiltonian}).
The canonical partition function $Z\left(\beta,B,J\right)$ is given by the
summation of $\exp\left(-\beta H_{s}\right)$ over all possible microstates $s$:
$Z\left(\beta,B,J\right)=\sum_{s}e^{-\beta H_{s}}$, where
$\beta=\left(k_{B}T\right)^{-1}$ is the inverse temperature.  In the
alternative expression of the reliability polynomial Eq.~(\ref{eq:RxrG}), the
summation over all subgraphs is organized into equivalence classes by the
number of edges in the subgraphs.  Similarly, we can group all microstates into
equivalence classes (energy levels) determined by the number of adjacent sites
in opposite states (``discordant vertex pairs" or ``edges'') and the number of
spin-up sites.  With $N\equiv\left|V\right|$, the partition function can be
expressed as: 
\begin{equation}
	Z\left(\beta,B,J\right)
    = C 
    \sum_{k=0}^{M}\sum_{v=0}^{N}
	g\left(k,v\right)e^{-2\beta\left(Jk+\mu Bv\right)}
	\label{eq:ZBT}
\end{equation} 
where $C\equiv e^{\beta\left(JM+\mu BN\right)}$ and $g\left(k,v\right)$ is the
number of microstates with $v$ spin-up vertices and $k$ discordant adjacent
vertex pairs (edges).  Note that in the absence of an external field ($B=0$),
the sum over $v$ reduces to the univariate density of states $g\left( k
\right)$.  Eq.~(\ref{eq:ZBT}) is a useful form for deriving a
``low-temperature'' expansion~\cite{Pathria1996Book}, in which only equivalence
classes with small $k$ and $v$ contribute.  In analogy with the reliability
polynomial Eq.~(\ref{eq:RxrG}), each term in Eq.~(\ref{eq:ZBT}) can be factored
into two separate parts: \textit{structural} -- the number of microstates
$g\left(k,v\right)$ determined by the graph, and \textit{dynamical} -- the
physical quantities $\beta$, $J$ and $B$-- or \textit{thermal} to be more
precise in the Ising model context.  Just as the structural factors $R_k(r, G)$
of the reliability $R\left( x; r, G \right)$ can be computed independently of
$x$, Eq.~\ref{eq:Rk}, $g\left( k,v \right)$ can be computed independently of
$\beta$, $B$ or $J$.  Once we have $g\left( k,v \right)$, we can plug in any
value of physical quantities and compute the thermodynamic functions without
any further simulation. This is more efficient than the traditional Metropolis
methods.  This observation has also been made by Wang and
Landau~\cite{Wang2001,Wang2001a}. By introducing the transformation
$x\left(B,\beta\right)\equiv\left(1+e^{2\beta\mu B}\right)^{-1}$ and
$y\left(J,\beta\right)\equiv\left(1+e^{2\beta J}\right)^{-1}$, we can express
the partition function $Z\left(\beta,B,J\right)$ as a bivariate reliability
polynomial $R\left(x,y;r,G\right)$ using the transformation $\beta\mu B
    \equiv \frac{1}{2}\ln\frac{1-x}{x}$ and $\beta J
    \equiv\frac{1}{2}\ln\frac{1-y}{y}$ (Appendix~\ref{apdx:A}):
\begin{eqnarray*} 
	Z\left(\beta,B,J\right) & \propto & 
	\sum_{v,k}g\left(k,v\right)x^{v}\left(1-x\right)^{N-v}y^{k}\left(1-y\right)^{M-k}\\
	& = & 
    R\left(x,y;r,G\right)
\end{eqnarray*} 
Note that the density of states $g\left( k,v \right)$ is equivalent to $R_k$,
the number of subgraphs satisfying a binary criterion, Eq.~\ref{eq:Rk}.  We
call the corresponding reliability criterion $r$ \textit{Ising-feasibility}: a
subgraph $s$ is Ising-feasible if and only if it is possible to find an
assignment of spins to all vertices such that every pair of discordant vertices
connected by an edge in $G$ is also connected by an edge in $s$ and there is no
edge between any other pair of vertices. Fig.~\ref{fig:IsFeasible}a illustrates
an Ising-feasible microstate on a 4-by-4 square lattice;
Fig.~\ref{fig:IsFeasible}b, an infeasible one.  Thus the Ising model's
partition function is a bivariate reliability polynomial with the special
Ising-feasibility criterion.

\section{$\delta-$Function Approximation}\label{sec:deltaAppr}

By definition, the structural factor $g\left( k,v \right)$ is independent of any of the
physical variables, $\beta$, $J$, or $B$.  Solving the Ising model
numerically on any graph $G$ requires only estimating its joint density of
states $g\left( k,v \right)$. 
Given the joint density of states, the partition function, and thus any
thermodynamic quantities, can easily be evaluated for any particular values of
 $\beta$, $J$ and $B$.  We use Monte Carlo sampling to estimate $g\left( k,v
\right)$.  Because sampling vertices and edges independently rarely produces an
Ising-feasible configuration, 
we randomly assign $v$ vertices to be in the spin-up state and then measure the
number of discordant node pairs (edges) $k$. 
We then estimate the conditional probability $p\left(k|v\right)$
by the frequency of producing $k$ edges given $v$ spin-up vertices. 
Because there are exactly ${N \choose v}$ ways to choose $v$ vertices, the
joint density of states $g\left(k,v\right)$ can be expressed as
$p\left(k|v\right){N \choose v}$.  For example, on a 2D square lattice with
periodic boundary conditions, 
when $v=1$ the only feasible microstates have $k = 4$. Therefore,
the number of microstates with 4 edges and 1 spin-up vertex is
$g\left(4,1\right)=1\cdot{N \choose 1}=N$.
Similarly, for $v=2$, $k=6$ when the two chosen vertices are neighbors and
$k=8$ when they are not. Assuming $N\geq4$, the corresponding
conditional probabilities $p\left(k=6|v=2\right)=\frac{4}{N-1}$
and $p\left(k=8|v=2\right)=\frac{N-5}{N-1}$.  The
number of states $g\left(k=6,v=2\right)$ and $g\left(k=8,v=2\right)$ can be
calculated accordingly by multiplying ${N \choose 2}$. 
These are the lowest order terms in the low temperature expansion.
In general, $p\left(k|v\right)$ is very difficult
to compute analytically.  

\begin{figure}
\includegraphics[width=\columnwidth]{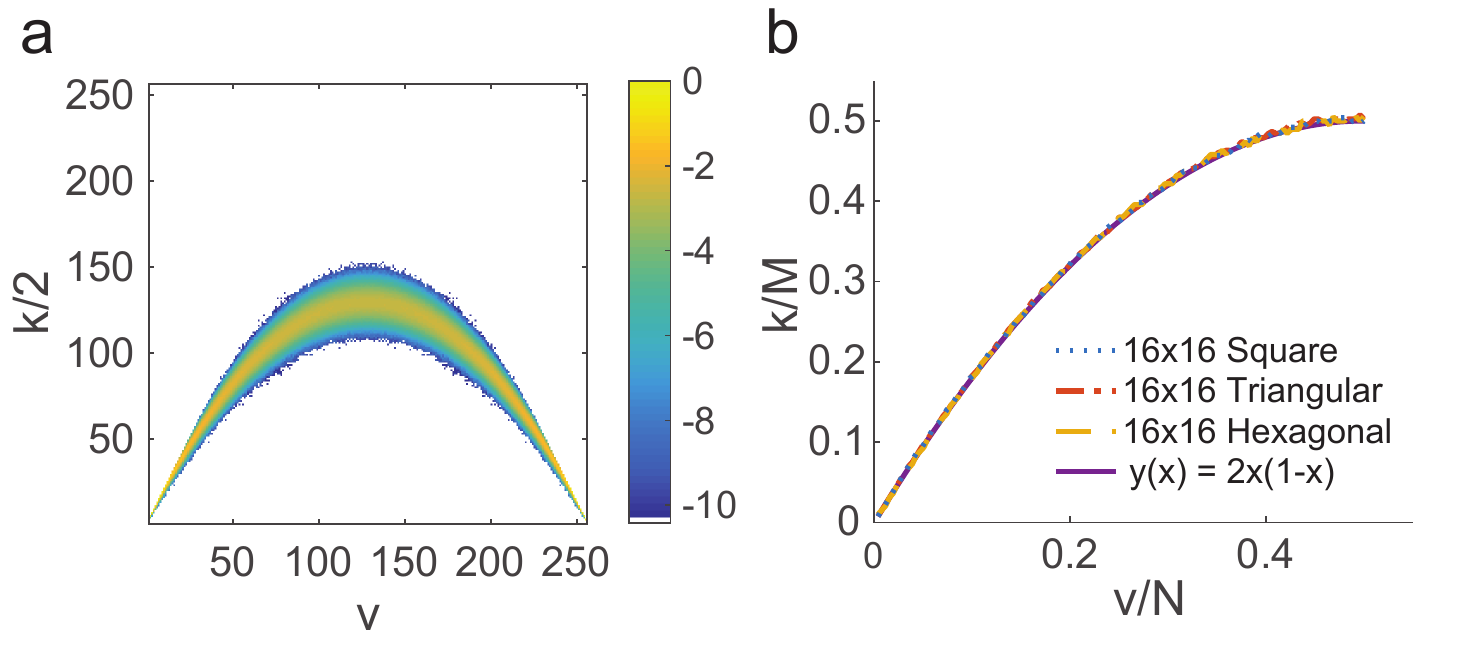}
\protect\caption{\label{fig:deltaApproxUandC}
(a) Conditional state distribution of $p\left(k|v\right)$ sampled by a na\"ive 
Monte Carlo simulation on a 16-by-16 square lattice. The conditional probability
is normalized separately for each value of $v$. The color reflects the value of $p\left(k|v\right)$ in 
logarithmic scale. As $p\left(k|v\right)$ is symmetric along $v=N/2$, the simulation 
is only done for $v\leq N/2$. 
(b) The peaks of $p\left(k|v\right)$ for various lattices have the same functional form:
$y\left(x\right)=2x\left(1-x\right)$, where $x\equiv v/N$ and $y\equiv k/M$. 
}
\end{figure}

An example of $p\left(k|v\right)$ sampled using a na\"ive
Monte Carlo method on a 16-by-16 square lattice is shown in
Fig.~\ref{fig:deltaApproxUandC}a. Note that, because $p\left( k|v \right)$ is
the \textit{conditional} density function, it is normalized separately for each value of $v$ so that $\sum_k p\left( k|v
\right) =1$.  Also, for a 16-by-16 square lattice, $N=256$ and
$M=512$, the maximum of $k$ can be as great as 512.
This maximum is only achieved by a microstate in which spin-up and spin-down
sites strictly alternate.  There are only two such states out of ${256 \choose
128}$ possible microstates with v=128.  The na\"ive Monte Carlo method described
above can hardly be expected to sample microstates as rare as this.
Interestingly, as we explain in Section~\ref{sec:joint}, these rare microstates
can dominate the value of the joint density of states.

Empirically, the peaks of $p\left(k|v\right)$ lie on the curve
$\frac{k}{M}=2\frac{v}{N}\left(1-\frac{v}{N}\right)$.  This functional
relationship seems independent of the system size $N$ or the coordination
number (mean degree) $q\equiv2M/N$ of the lattice,
Fig.~\ref{fig:deltaApproxUandC}b.  A simple argument suggests why this is the case.
If the spin-up vertices are distributed uniformly across the lattice, 
the probability that the neighbor of a spin-up vertex is spin-{\em down} is $1 - \frac{v}{N}$.
For $v$ spin-up vertices, each with $q$ neighbors, the expected number of discordant pairs is
thus $\frac{1}{2}qv(1 - \frac{v}{N})$.

As the system size goes to infinity, $N\rightarrow\infty$, the conditional
probability $p\left(k|v\right)$ becomes more sharply peaked at its center. We
can approximate $p\left(k|v\right)$ as a Kronecker $\delta$-function
$p\left(k|v\right)\simeq\delta\left(\frac{k}{M},2\frac{v}{N}\left(1-\frac{v}{N}\right)\right)$.
Inserting the $\delta$-function approximation for $p\left(k|v\right)$ in our expression for the partition function, Eq.~(\ref{eq:ZBT}), yields:
\begin{eqnarray}
    Z\left(\zeta,\eta\right) & = & C\sum_{e,v}p\left(k|v\right){N \choose v}e^{-2\left(\zeta v+\eta k\right)}\notag\\
    & \simeq & C\sum_{e,v}\delta\left(k/M,y\left(v/N\right)\right){N \choose v}e^{-2\left(\zeta v+\eta k\right)}\notag\\
    & \simeq & C\sum_{v}{N \choose v}e^{-2N\left(\zeta\frac{v}{N}+\frac{1}{2}\eta qy\left(v/N\right)\right)}
    \label{eq:Zintegral}
\end{eqnarray}
where $\zeta\equiv\beta\mu B$, $\eta\equiv\beta J$ and $y\left( x
\right)=2x\left( 1-x \right)$.  
This produces the Bragg-Williams mean-field approximation \cite{Bragg1934,Bragg1935}
    , where the interaction
term in the Hamiltonian $-J\sum_{\left(i,j\right)\in E}\sigma_{i}\sigma_{j}$ is
approximated as
$-J\left(\frac{1}{2}q\overline{\sigma}\right)\sum_{i}\sigma_{i}$, and
$\overline{\sigma}=\frac{1}{N}\sum_{i}\sigma_{i}$ is the average spin of the system.  

The Bragg-Williams mean-field approach -- and hence Eq.~\ref{eq:Zintegral} --
incorrectly predicts that one-dimensional systems exhibit a critical point.
According to Eq.~\ref{eq:Zintegral}, the partition function depends on the
dimension of the system and the graph structure only through $q$, the
coordination number, where $q=2$ for a 1D lattice,  $q=4$ for a 2D square
lattice and $q=6$ for a 2D triangular lattice.  Moreover, its dependence on $q$
is only through the product $N\eta qy(x)$.  If the external field is zero
($\zeta=0$), changing $q$ is equivalent to changing the system size $N$ or
coupling strength $\eta$.  In other words, a 2D square lattice with size $N$
behaves the same as a 1D lattice with size $2N$ in this approximation, which is
physically incorrect.  In Section~\ref{sec:joint} we explore the causes of this
failure and explain how to address it.

\section{Estimating the Density of States}\label{sec:joint} 

\begin{figure}
\centering\includegraphics[width=0.6\columnwidth]{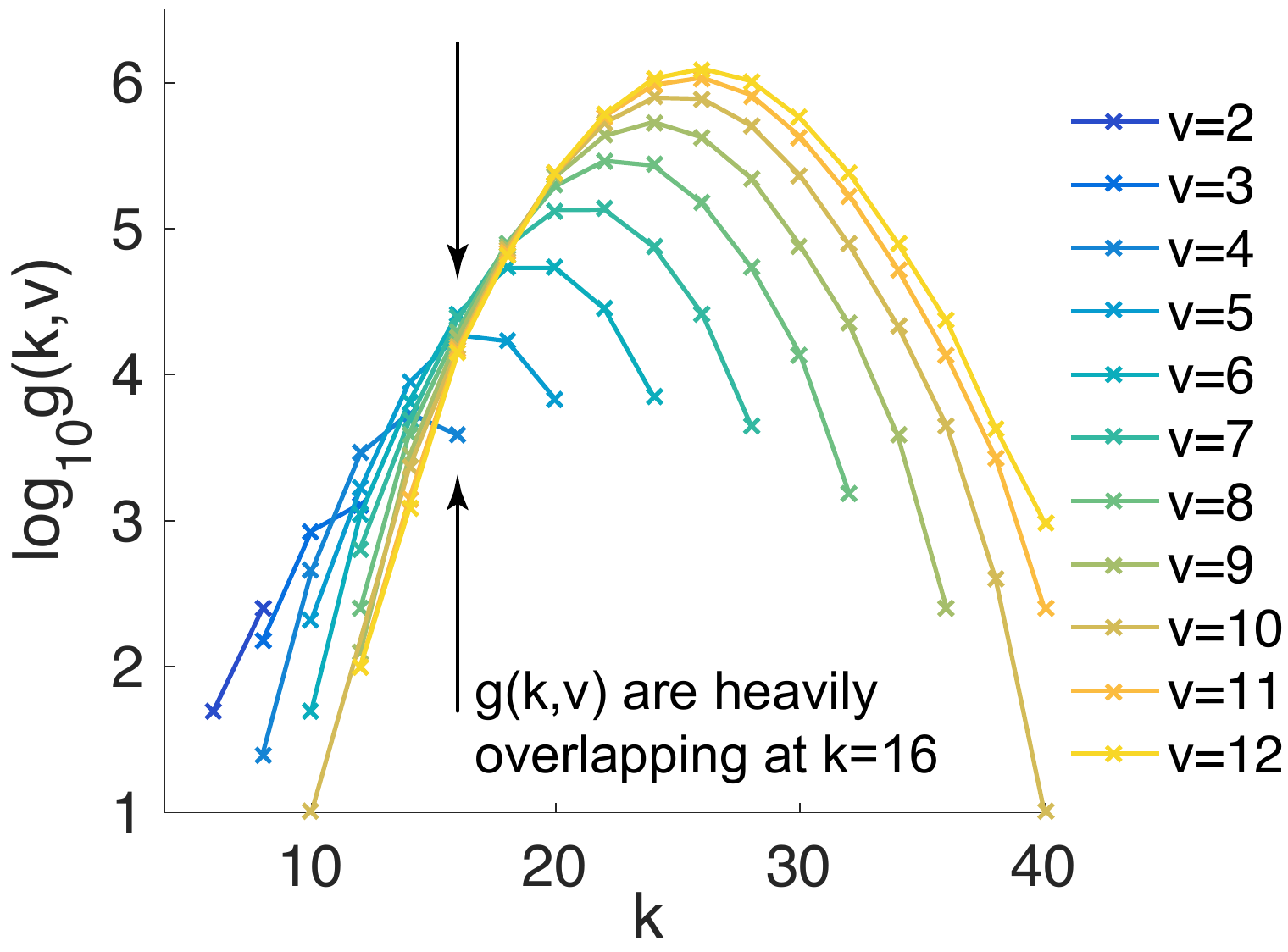}
\protect\caption{\label{fig:gkv5by5} 
The exact joint density of states $g\left(k,v\right)$ computed via exhaustive
 enumeration on a 5-by-5 square lattice.  Because a na\"ive Monte Carlo only
samples points near the \textit{peak} of each curve, and the tails of many curves are
as important as the peaks of others, it severely underestimates the univariate density of states $g\left( k
\right) \equiv \sum_v g\left(k,v\right)$ at $k=16$.
}
\end{figure}

Although, for a \textit{particular} $v$, it is reasonable to approximate
$p\left(k|v\right)$ as a $\delta$-function, critical phenomena are determined
by  \textit{all} $p\left(k|v\right)$ synergistically.  The Ising model is hard
to solve exactly because extremely rare events for one value of $v$ are as
important as the most common events for another value.  To demonstrate this, we
first transform the conditional probability $p\left(k|v\right)$ to the number
of states $g\left(k,v\right) = p\left(k|v\right){N \choose v}$.  Since the
binomial factor ${N \choose v}$ scales exponentially with $v$, it can dominate
the ratio $g\left(k,v_i\right) / g\left(k,v_j\right)$.  \textit{Ceteris
paribus}, this makes contributions to $Z$ from the tails of $p(k|v_i)$
comparable to contributions from the peaks of $p\left(k|v_j\right)$.  The joint
density of states of a 5-by-5 2D lattice is shown in Fig.~\ref{fig:gkv5by5}.
Consider $g\left(16,5\right)$, the number of microstates with $k=16$ discordant
neighbors when there are $v=5$ spins up.  It corresponds to the \textit{peak}
of $p\left(k|5\right)$, and is roughly the same as $g(16,10)$, which is in the
\textit{tail} of $p\left(k|10\right)$.  The na\"ive Monte Carlo method misses
the tail of $p\left(k|v\right)$, and is thus inaccurate.

Despite the failure of the na\"ive Monte Carlo method, the strategy of dividing
energy states into equivalence classes remains valuable. It separates the
estimation of the joint density of states $g\left(k,v\right)$ into $N/2$
independent estimations of univariate distributions $p\left(k|v\right)$, thus
enabling a novel parallel estimation scheme. And, each of $p\left(k|v\right)$
can be estimated using the improved Wang-Landau (WL)
algorithm\cite{Wang2001,Wang2001a}. The WL algorithm is a Markov-chain Monte
Carlo algorithm to obtain the univariate density of states $g\left(k\right)$
for the Ising model.

The WL algorithm is very similar to the
Metropolis-Hasting~\cite{Metropolis1953,HASTINGS1970} algorithm.  However,
instead of \textit{assuming} the detailed balance condition, the WL algorithm
pursues its so-called ``flat'' histogram by sculpting the $g\left(k\right)$
gradually during the simulation.  Therefore, the running time of the WL
algorithm largely depends on the number of energy states. As the number of
states in $g\left(k,v\right)$ is proportional to $O\left( N^2 \right)$, the
\textit{square} of the number of states in $g\left( k \right)$, the WL
algorithm takes a tremendous amount of time to converge when computing the
joint density of states~\cite{landau2004}.  Each step in the random walk in WL
algorithm flips the spin of a random vertex, which inevitably changes both $v$
and $k$. Our modification of this algorithm is to constrain the random walk to
maintain $v$ invariant.  For each $v$-spin subspace, we assign an independent
random walker.  Therefore, the number of energy states is reduced to $O\left( N
\right)$ for each walker.  Specifically, instead of randomly flipping the spin
of a vertex as is done in the WL random walk, each step of our random walk
exchanges the locations of a spin-up vertex and spin-down vertex. The rest of
the algorithm is as the same as the WL algorithm \cite{Wang2001},
Appendix~\ref{apdx:C}. 

\begin{figure}
\centering\includegraphics[width=0.6\columnwidth]{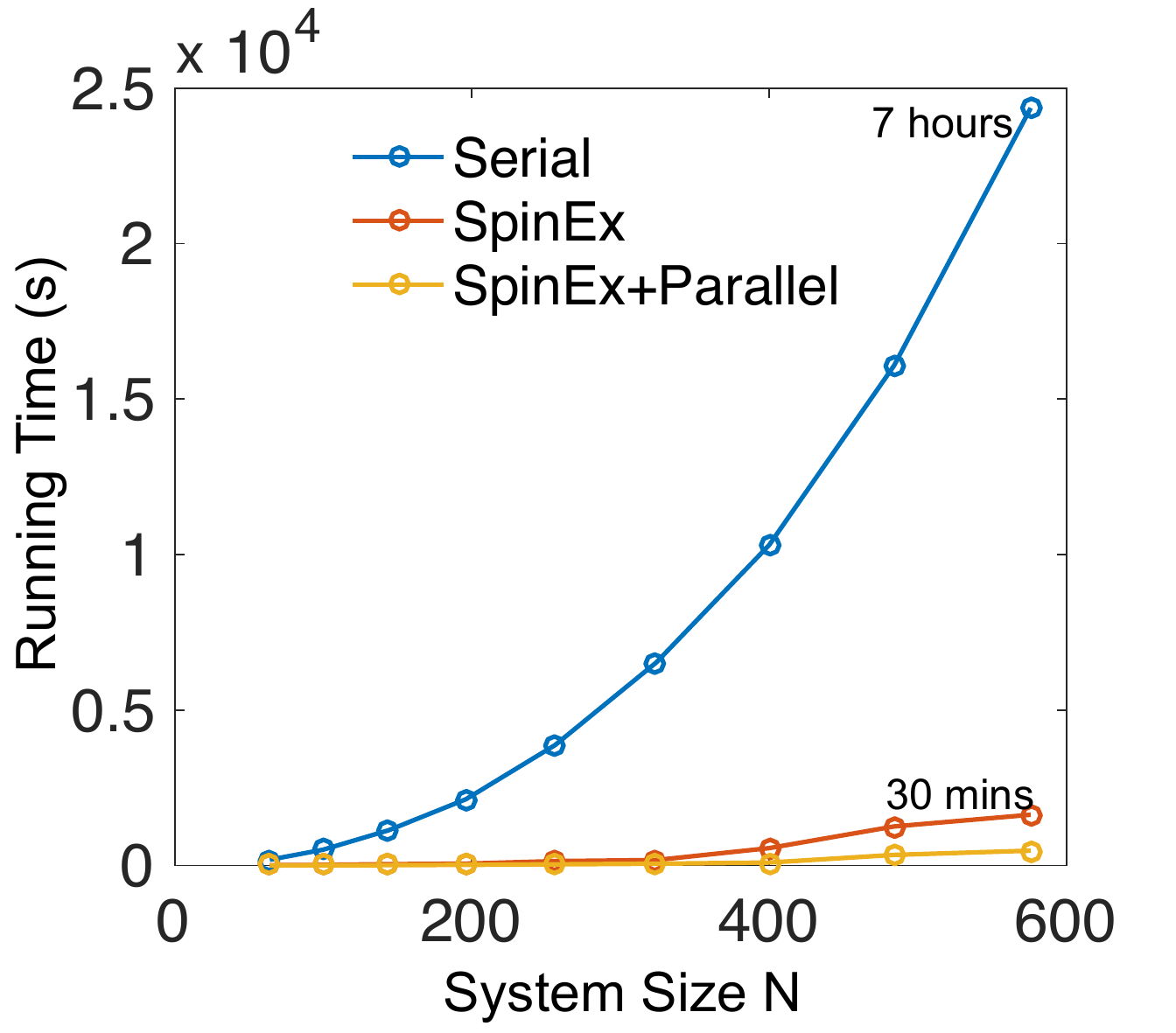}
\protect\caption{\label{fig:Time} 
The running time for estimating the joint density of states $g\left( k,v \right)$
on 2D lattices of different sizes, from $N=8\times 8$ to $N=24\times 24$.
The sequential Wang-Landau algorithm (blue) needs to cover $O\left( N^2 \right)$ energy states,
and become impractical for large systems.
The spin-exchange WL algorithm (red) divides the energy states into $N/2$ energy slices,
and the running time is bounded by the energy slice with the most number of states at $v=N/2$.
This energy slice only contains $O\left( N \right)$ states.
By dividing the energy slice into 6 equal-sized, 75\% overlapping energy windows,
the running time is reduced even further (yellow). 
}
\end{figure}

To demonstrate the efficiency of our algorithm, we compare the running
time on 2D lattices of different sizes, from $N=8\times 8$ to $N=24\times
24$, Fig.~\ref{fig:Time}. The number of energy states is proportional to $N^2$.
The running time for the sequential WL algorithm (blue) grows exponentially as the 
number of energy states increases. It becomes impractical for large systems $N>10^3$. 
The spin-exchange method (red) splits the energy states into $N/2$ 
$v$-specific energy slices of different sizes. 
The overall running time is bounded by that of the slice for which $v=N/2$, which contains the most energy states.
As the number of energy states in each slice is $O\left( N \right)$, the spin-exchange
method is much faster than the sequential WL algorithm.
Since each energy slice essentially is a univariate density, we can reduce the
computation time even further by dividing an energy slice into multiple overlapping energy windows.
We tested using six $75\%$ overlapping energy windows (yellow).
The running time test simply assumes independent random walkers in each window.
One can choose more sophisticated methods, 
such as the replica-exchange scheme~\cite{Vogel2013,Vogel2014}.

\begin{figure}
\centering\includegraphics[width=\columnwidth]{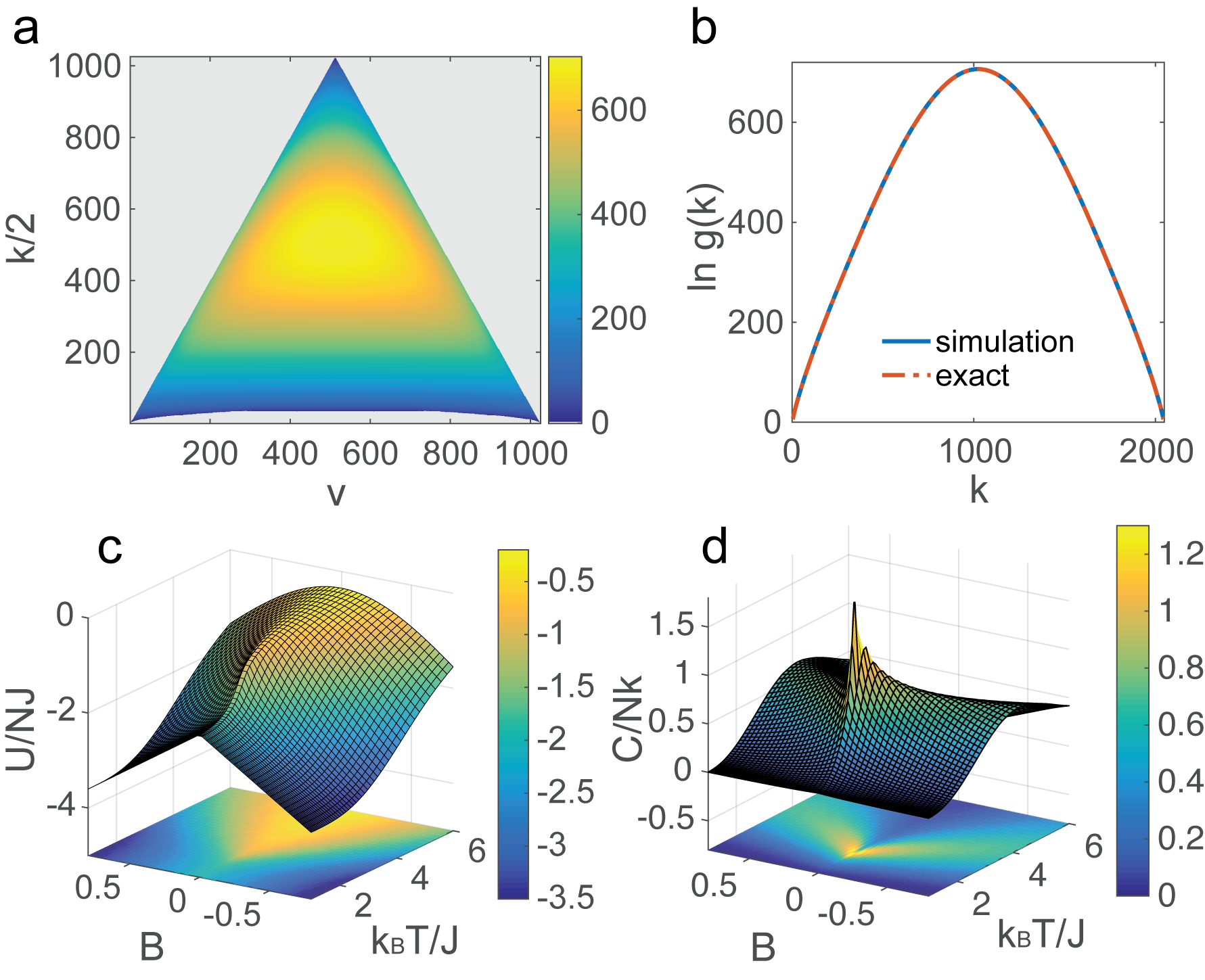}
\protect\caption{\label{fig:UC2D} 
(a) The joint density of states $g\left( k,v \right)$ of a 32-by-32 square lattice.
(b) The univariate density of states $g\left( k \right)\equiv \sum_v g\left(
k,v \right)$ estimated using a spin-exchange MCMC algorithm,
compared with the exact analytical result.
(c) The internal energy $U=-\partial \ln Z / \partial \beta$, and
(d) the heat capacity $C=\partial U / \partial T$
with coupling strength $J=0.5$. }
\end{figure}

We apply our algorithm on a 32-by-32 square lattice, with $\sim 0.5\times 10^6$
energy levels (equivalence classes) and more than $10^{308}$ microstates. Note
that, for the same system, the univariate density of states $g\left( k \right)$
only has $\sim 10^3$ energy levels .  Fig.~\ref{fig:UC2D}a shows estimates for
the joint density of states.  The simulation is performed using 300 cores
within two days.  To verify this result, we compare its projection onto the
univariate density of state $g\left( k \right)\equiv\sum_v g\left( k,v \right)$
with the known analytical result~\cite{beale1996}. As shown in
Fig.~\ref{fig:UC2D}b, the agreement is very good.  Given the joint density of
states, we can easily find the partition function $Z\left( \beta,B,J \right)$
using Eq.~(\ref{eq:ZBT}).  Then, without additional simulations, any
thermodynamic functions can be obtained directly from the partition function,
such as the internal energy $U=-\partial \ln Z / \partial \beta$ and the heat
capacity $C=\partial U / \partial T$.  Fig.~\ref{fig:UC2D}c and d show the
internal energy $U$ and heat capacity $C$ as a function of inverse temperature
$\beta$ and external field $B$ (assuming the magnetic susceptibility $\mu=1$)
at $J=0.5$.  The heat capacity curve presents the correct critical point of
$k_BT/J=2.27$ at $B=0$. As the heat capacity is known to be very sensitive to
the density of states, in Fig.~\ref{fig:ApdxFig}a Appendix~\ref{apdx:D}, we
show that the heat capacity at $B=0$ from our simulation agrees with the one
from the analytic result. 

\begin{figure}
\centering\includegraphics[width=\columnwidth]{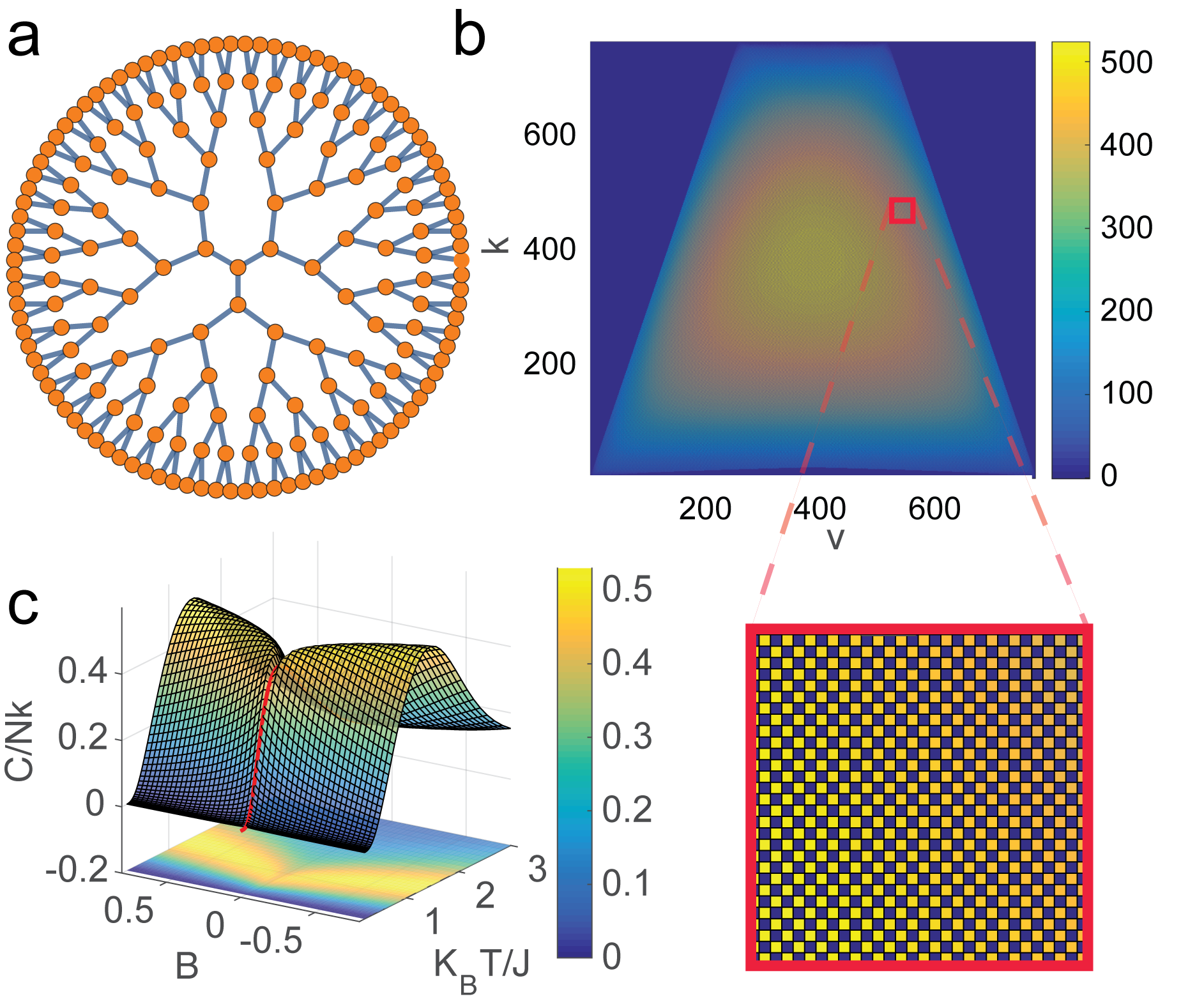}
\protect\caption{\label{fig:Cayley} 
(a) A Cayley tree with degree $d=3$ and number of shells $r=6$.
(b) The joint density of states $g\left( k,v \right)$ of the Cayley tree with $d=3$ and $r=8$ ($N=765$).
    As $d=3$, there are many inaccessible states.
(c) The heat capacity of the Ising model on the Cayley tree. The red curve is from the exact solution at $B=0$.
}
\end{figure}

We also apply our algorithm on {\cg Cayley trees (a finite-size analogue to Bethe lattices)},
where the exact result of the Ising model for $B=0$ is
known~\cite{baxter1982,eggarter1974}.  A Cayley tree has a central vertex and
every vertex (except leaves) has $d$ neighbors, Fig.~\ref{fig:Cayley}a.  It is
defined by two parameters, the degree $d$ and the number of shells $r$.  There
are $d \left( d-1 \right)^{\left( j-1 \right)}$ vertices at $j$-th shell and
$d\left[ \left( d-1 \right)^r -1 \right]/\left( d-2 \right)$ vertices in total.
So the ratio of the number of leaves to the system size tends to $\left( d-2
\right)/\left( d-1 \right)$. The dimensionality
$\lim_{n\rightarrow\infty}\left( \ln c_n \right)/\ln n \rightarrow \infty$,
where $c_n$ is the number of vertices within $n$ shells. All these
characteristics make Cayley trees very different from a regular lattice and
very interesting to study. The simulation on a Cayley tree with $d=3$ and $r=8$
($N=765$) yields the joint density of states as shown in
Fig.~\ref{fig:Cayley}b. As $d=3$ in this particular Cayley tree, there are
inaccessible states (``holes'') in the $g\left( k,v \right)$.  The heat
capacity is shown in Fig.~\ref{fig:Cayley}c and a comparison with the analytic
result at $B=0$ is shown in Fig.~\ref{fig:ApdxFig}b in Appendix~\ref{apdx:D}.  
{\cg Due to the difference in topologies, the heat capacity of the Ising model
on the Cayley tree is very different from that on the 2D square lattice.}

The spin-exchange WL algorithm proposed above provides a unique and efficient
parallel scheme for computing the joint density of states of Ising models in
the presence of an external field. Essentially, this parallel scheme splits the
joint density of states $g\left(k,v \right)$ into $N/2$ conditional densities
$p\left( k|v \right)$.

\section{Conclusion}
Network reliability is a general framework for understanding the interplay of
network topology and network dynamics. Here we have used network reliability to
study a prototypical network dynamical system -- the Ising model. This
framework can be adapted to other network dynamics as well, by defining a
suitable feasibility criterion for microstates. 

The network reliability perspective separates effects of network structure from
dynamics in the system's partition function.  Based on this separation, we
introduced a $\delta$-function approximation for the density of states, which
leads to the Bragg-Williams approximation for the internal energy. We also
showed why a na\"ive Monte Carlo method is not accurate enough for estimating
the joint density of states.  Finally, we introduced a novel parallel scheme
using a spin-exchange MCMC algorithm for estimating the joint density of
states.  The scheme requires no inter-processor communication and can take
advantage of the replica-exchange parallel framework. We applied our method to
a periodic 32-by-32 square lattice estimating its internal energy and heat
capacity as a function of both temperature and external magnetic field. 

This work will make simulations of Ising-like dynamics on large, complex
networks feasible and efficient, and opens the door to studying the Ising model
in the presence of an external field.
An efficient algorithm makes it possible to study the effects of network structure
in systems that are too irregular to admit closed-form solutions.  Furthermore,
as is suggested by Fig.~\ref{fig:UC2D}d, the nature of the phase transition
depends on the external field strength.  Our approach enables studies of such
phenomena in large systems for the first time.

\begin{acknowledgments} 
    Research reported in this publication was supported by
    the National Institute of General Medical Sciences of the National
    Institutes of Health under Models of Infectious Disease Agent Study Grant
    5U01GM070694-11, by the Defense Threat Reduction Agency under Grant
    HDTRA1-11-1-0016 and by the National Science Foundation under Network
    Science and Engineering Grant CNS-1011769. The content is solely the
    responsibility of the authors and does not necessarily represent the
    official views of the National Institutes of Health, the Department of
    Defense or the National Science Foundation. We would like to 
    thank P.  D. Beale for providing the code to compute exact univariate 
    density of states, and T. Vogel for discussing the replica-exchange 
    algorithm. We would also like to thank Y. Khorramzadeh, Z. Toroczkai, 
    M. Pleimling, U. T\"{a}uber and R. Zia for comments and suggestions.
\end{acknowledgments}

\appendix
\section{\label{apdx:A}} 
Here we show that the Ising model's partition function, Eq.~(\ref{eq:ZBT}), can
be expressed as the network reliability Eq.~(\ref{eq:RxrG}) under the
transformation $x\left( \zeta 
\right)\equiv\left( 1+e^{2\zeta} \right)^{-1}$ and $y\left( \eta 
\right)\equiv \left( 1+e^{ 2 \eta } \right)^{-1}$, where $\zeta \equiv \beta \mu B$ and
$\eta \equiv \beta J$. The inverse transformations 
are $\zeta \equiv \frac{1}{2}\ln\frac{1-x}{x}$ and $\eta\equiv\frac{1}{2}\ln\frac{1-y}{y}$.
Plugging $\zeta$ and $\eta$ into the partition function Eq.~(\ref{eq:ZBT}) gives:

\begin{widetext}

\begin{eqnarray*}
Z\left( \zeta,\eta \right) &=& e^{\eta M + \zeta N} 
\sum_{k=0}^M \sum_{v=0}^N g\left( v,k \right) e^{-2\eta k -2 \zeta v} \\
&=& e^{\left( \frac{1}{2} M \ln \frac{1-y}{y} + \frac{1}{2} N \ln \frac{1-x}{x} \right)}
\sum_{k=0}^M \sum_{v=0}^N g\left( v,k \right) e^{-k\ln \frac{1-y}{y}  - v\ln \frac{1-x}{x} } \\
&=& \left( \frac{1-y}{y} \right)^{\frac{M}{2}} \left( \frac{1-x}{x} \right)^{\frac{N}{2}}
\sum_{k=0}^M \sum_{v=0}^N g\left( v,k \right)  \left(\frac{y}{1-y}\right)^k 
\left(\frac{x}{1-x}\right)^v \\
&=& \left( y\left( 1-y \right) \right)^{-\frac{M}{2}}\left( x\left( 1-x \right) \right)^{-\frac{N}{2}}
\sum_{k=0}^M \sum_{v=0}^N g\left( v,k \right) y^k \left( 1-y \right)^{M-k} x^v\left( 1-x \right)^{N-v} \\
&=& \left( y\left( 1-y \right) \right)^{-\frac{M}{2}}\left( x\left( 1-x \right) \right)^{-\frac{N}{2}}
R\left( v,k;r,G \right)
\end{eqnarray*}

\end{widetext}
\section{\label{apdx:C}} 
The spin-exchange WL algorithm starts with a choice of $v\in[2,N/2]$, a prior
unknown $p\left(k|v\right)$ and a histogram $H\left(k|v\right)$. At each step of the random
walk, the system selects a new state $k'$ with probability 
$P=\min\left( 1, p\left(k|v\right) / p\left(k'|v\right) \right)$.
The $p\left(k^* | v\right)$ and $H\left(k^* |v\right)$ of the accepted state
$k^*$ ($k^*=k$ or $k'$) will be updated: 
$p\left(k^* | v\right) \leftarrow f\cdot p\left(k^* | v\right) $ and 
$H\left(k|v\right)\leftarrow H\left(k|v\right) + 1 $, where $f$ is a modification factor. Once the histogram
$H\left(k|v\right)\geq 1/\sqrt{f},\ \forall k$ is sufficiently
``flat''\cite{Zhou2005}, it is reset to zero
$H\left(\cdot|v\right)\leftarrow 0$, and the modification factor is downscaled
$f\leftarrow \sqrt{f}$.  The simulation stops when the modification
factor $f$ is very close to 1, e.g. $\exp\left( 10^{-6} \right)$.

\section{\label{apdx:D}}
\begin{figure}[h]
\centering\includegraphics[width=0.8\columnwidth]{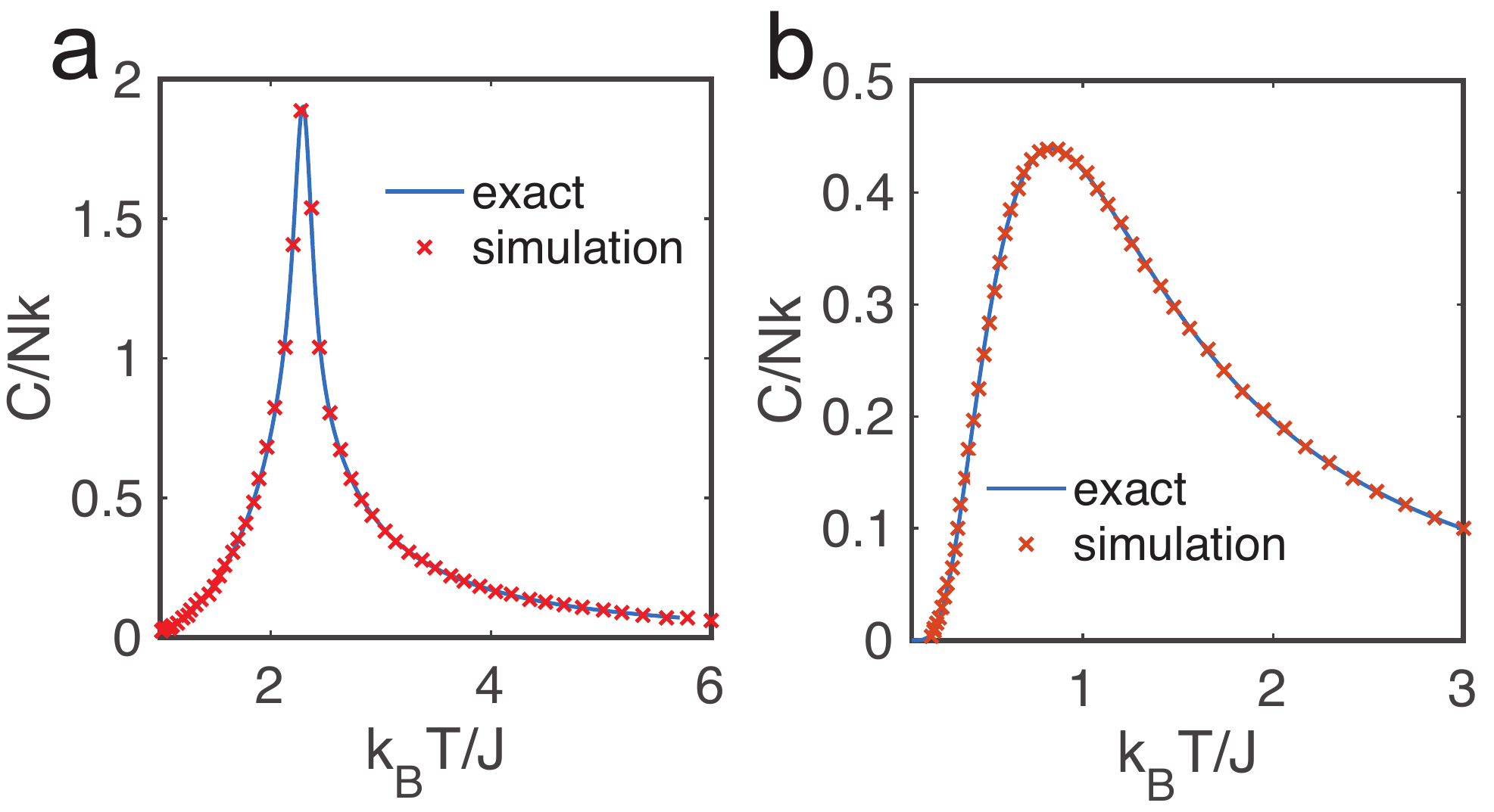}
\protect\caption{\label{fig:ApdxFig} 
Compare the heat capacity with the known results at $B=0$ (without external field).
(a) On a $32\times 32$ 2D square lattice. The exact solution is calculated from the 
exact density of states $g\left( k \right)$ from P.D. Beale~\cite{beale1996}.
(b) On a $d=3$ and $r=6$ Cayley tree. The exact solution is computed as $C/k = -\beta^2 d^2\ln Z / d\beta^2$,
where $\ln Z$ is given by~\cite{eggarter1974}.
}
\end{figure}

%

\end{document}